\documentclass[a4paper,11pt]{article}
\usepackage{jcappub}

\def\a{\widetilde{\alpha}}
\def\b{\widetilde{\beta}}
\def\m{\overline{m}}

\newcommand{\rem}[1]{}

\begin{document}
\title{The portrait of eikonal instability in Lovelock theories}

\author[\dagger\star]{R. A. Konoplya,}
\emailAdd{roman.konoplya@gmail.com}
\affiliation[\dagger]{Theoretical Astrophysics, Eberhard-Karls University of T\"ubingen, T\"ubingen 72076, Germany}
\affiliation[\star]{Institute  of  Physics  and  Research  Centre  of  Theoretical  Physics  and  Astrophysics, Faculty  of  Philosophy  and  Science,  Silesian  University  in  Opava,  Opava,  Czech  Republic}

\author[\ddagger]{A. Zhidenko}
\emailAdd{olexandr.zhydenko@ufabc.edu.br}
\affiliation[\ddagger]{Centro de Matem\'atica, Computa\c{c}\~ao e Cogni\c{c}\~ao, Universidade Federal do ABC (UFABC),
Rua Aboli\c{c}\~ao, CEP: 09210-180, Santo Andr\'e, SP, Brazil}

\abstract{
Perturbations and eikonal instabilities of black holes and branes in the Einstein-Gauss-Bonnet theory and its Lovelock generalization were considered in the literature for several particular cases, where the asymptotic conditions (flat, dS, AdS), the number of spacetime dimensions $D$, non-vanishing coupling constants ($\alpha_1$, $\alpha_2$, $\alpha_3$ etc.) and other parameters have been chosen in a specific way. Here we give a comprehensive analysis of the eikonal instabilities of black holes and branes for the \emph{most general} Lovelock theory, not limited by any of the above cases. Although the part of the stability analysis is performed here purely analytically and formulated in terms of the inequalities for the black hole parameters, the most general case is treated numerically and the accurate regions of instabilities are presented. The shared Mathematica\textregistered{} code allows the reader to construct the regions of eikonal instability for any desired values of the parameters.
}

\maketitle

\section{Introduction and a review of the literature on black-hole perturbations in Lovelock theories}

Deviations or corrections to the Eintein theory of gravity emanate from a number of fundamental problems in the modern theoretical physics, such as hierarchy problem, dark energy/dark matter problem, attempts to construct quantum gravity and resolve the singularity problem, etc. Current observations of black holes in the gravitational and electromagnetic spectra do not close an opportunity for alternative theories \cite{TheLIGOScientific:2016src,Konoplya:2016pmh}. Theories with higher curvature corrections to the Einstein action are important in various areas of astrophysics and high energy physics. The Lovelock theory of gravity \cite{Lovelock:1971yv} is the most general mathematically consistent metric theory leading to second order equations of motion in arbitrary number of spacetime dimensions $D$. Thus, it is natural generalization of Einstein's gravity, coinciding with the Einsteinian equations of motion in $D=4$ world, but different for higher $D$. Asymptotically flat (or de Sitter) solutions in Einstein-Lovelock theory may represent black holes in higher dimensional gravity \cite{Boulware:1985wk}, allowing for the quantum corrections. This leads to a number of interesting consequences, such as, e.g., strong suppression of the intensity of Hawking evaporation even at geometrically moderate higher curvature corrections \cite{Rizzo:2006uz}. Asymptotically anti-de Sitter solutions are important backgrounds for analysis of intermediate t'Hooft coupling regime  \cite{Grozdanov:2016vgg,Brigante:2007nu}  within the AdS/CFT correspondence.

The Lagrangian of the Einstein-Lovelock theory has the form
\cite{Lovelock:1971yv}:
\begin{eqnarray}
  \mathcal{L} &=& -2\Lambda+\sum_{m=2}^{\m}\frac{\alpha_m}{m}\mathcal{L}_m,\label{gbg1} \\\nonumber
  \mathcal{L}_m &=& \frac{1}{2^m}\delta^{\mu_1\nu_1 \ldots\mu_m\nu_m}_{\lambda_1\sigma_1\ldots\lambda_m\sigma_m}\,R_{\mu_1\nu_1}^{\phantom{\mu_1\nu_1}\lambda_1\sigma_1} \ldots R_{\mu_m\nu_m}^{\phantom{\mu_m\nu_m}\lambda_m\sigma_m},\label{gbg2}
\end{eqnarray}
where
$$\delta^{\mu_1\nu_1\ldots\mu_m\nu_m}_{\lambda_1\sigma_1\ldots\lambda_m\sigma_m}=\det\left(
\begin{array}{cccc}
\delta^{\mu_1}_{\lambda_1} & \delta^{\mu_1}_{\sigma_1} & \cdots & \delta^{\mu_1}_{\sigma_m} \\
\delta^{\nu_1}_{\lambda_1} & \delta^{\nu_1}_{\sigma_1} & \cdots & \delta^{\nu_1}_{\sigma_m} \\
\vdots & \vdots & \ddots & \vdots \\
\delta^{\nu_m}_{\lambda_1} & \delta^{\nu_m}_{\sigma_1} & \cdots & \delta^{\nu_m}_{\sigma_m}
\end{array}
\right)$$
is the generalized totally antisymmetric Kronecker delta, which has no nonzero components for $m>\m=\left[{(D-1)/2}\right]$, $R_{\mu\nu}^{\phantom{{\mu\nu}}\lambda\sigma}$ is the Riemann tensor, $\alpha_1=1/16\pi G=1$ and $\alpha_2,\alpha_3,\alpha_4,\ldots$ are arbitrary constants of the theory.

The equations of motions following from the above Lagrangian say that the second order in curvature Gauss-Bonnet theory is the most general in $D=5$ and $6$, while the forth order Lovelock theory is the most general in $D=7, 8$, etc. Thus, the most general theory is too much involved and has a great number of parameters which represent qualitatively different situations. The number of spacetime dimensions $D$ implies the particular number of the terms of expansion. The cosmological constant $\Lambda$ can represent either asymptotically de Sitter ($\Lambda >0$) or anti-de Sitter ($\Lambda <0$) spacetimes which lead to qualitatively different boundary conditions and spectra. The black hole radius also describes essentially different objects: from small quantum black holes to very large ones, which can be approximated by planar horizons (black branes). Therefore, by now there are a number of papers devoted to perturbations and (in)stability of black holes in Lovelock theory, each of which is limited by looking  only at a particular case of the Lovelock theory. Here, we shall review the existing literature on this topic and propose the most general consideration of the eikonal instability for black holes and branes in the Lovelock theory (see Table~\ref{Table1}).

\begin{table}
\centering
\begin{tabular}{|c|c|c|c|c|}
  \hline
  type of black hole/brane & test fields & tensor-type & vector-type & scalar-type \\
   \hline
  GB BH & \cite{Konoplya:2004xx,Abdalla:2005hu,Zhidenko:2008fp,Gonzalez:2017gwa} & \cite{Dotti:2005sq,Konoplya:2008ix} & \multicolumn{2}{c|}{\cite{Gleiser:2005ra,Konoplya:2008ix}}\\
    \hline
  Lovelock BH & - & \cite{Takahashi:2010ye,Takahashi:2010gz,Yoshida:2015vua} & \multicolumn{2}{c|}{\cite{Takahashi:2010ye,Takahashi:2010gz,Yoshida:2015vua}}\\
    \hline
  Lovelock BH with charge & - & \cite{Takahashi:2011qda} & \multicolumn{2}{c|}{\cite{Takahashi:2012np}}\\
    \hline
  pure Lovelock BH & - & \cite{Gannouji:2013eka} & \multicolumn{2}{c|}{-} \\
    \hline
  GB-AdS BB & - & \multicolumn{3}{c|}{\cite{Konoplya:2017ymp,Grozdanov:2016vgg}} \\
    \hline
  GB-AdS BH & \cite{Abdalla:2005hu} & \multicolumn{3}{c|}{\cite{Konoplya:2017ymp}} \\
    \hline
  GB-dS BH & \cite{Abdalla:2005hu} & \multicolumn{3}{c|}{\cite{Cuyubamba:2016cug,Konoplya:2017ymp}} \\
    \hline
  Lovelock-AdS BB & - & \multicolumn{3}{c|}{\cite{Takahashi:2011du} ($D=10,11$)} \\
  \hline
  Lovelock-AdS BH & - &\multicolumn{3}{c|}{-} \\
  \hline
  Lovelock-dS BH & - &\multicolumn{3}{c|}{-} \\
  \hline
\end{tabular}
\caption{Review of papers on perturbations, quasinormal modes and stability of black holes and branes in the Lovelock theory.}\label{Table1}
\end{table}

Perturbations and quasinormal modes of Gauss-Bonnet black holes first were analyzed for the test scalar field  in the asymptotically flat \cite{Konoplya:2004xx}, asymptotically de-Sitter (dS) and anti-de Sitter (AdS) backgrounds \cite{Abdalla:2005hu}. Test fields analyzed by the time-domain integration \cite{Abdalla:2005hu,Zhidenko:2008fp,Gonzalez:2017gwa} showed no instability. On the contrary, it was found that the gravitational perturbations of the asymptotically flat Gauss-Bonnet black holes must be unstable and similar instability must occur also for asymptotically AdS and dS spacetimes \cite{Dotti:2005sq,Gleiser:2005ra}, though regions of instability were not presented in \cite{Dotti:2005sq,Gleiser:2005ra} for asymptotically non-flat cases. It turned out that the above instability has quite a remarkable behavior from the point of view of its spectra  \cite{Konoplya:2008ix}: the instability, counter-intuitively develops at high multipole numbers, while the lowest multipoles are stable. Later it was understood that such behavior was the consequence of the non-hyperbolicity of the master perturbation equations \cite{Reall:2014pwa} in the instability region: the absence of well-posed initial value problem showed up in the absence of convergence over various multipole numbers $\ell$ \cite{Konoplya:2008ix}. Since the instability is ``driven'' by long wavelengthes, it was called \emph{the eikonal instability} \cite{Cuyubamba:2016cug}.

Similar instability was found for the more general Lovelock theory \cite{Takahashi:2010ye,Takahashi:2010gz}, and the conclusion was that all sufficiently small black holes are unstable. This result was further extended to the case of charged black holes \cite{Takahashi:2011qda,Takahashi:2012np} and to the purely Lovelock (without the Einstein term) theory \cite{Gannouji:2013eka}. Still, while the idea of instability of small black holes was evident, the parametric region of the instability was not determined in the above papers. The parametric region of instability of Einstein-Gauss-Bonnet-de Sitter black holes was found in \cite{Cuyubamba:2016cug}, where it was shown that in addition to the eikonal instability there is a non-eikonal instability at the lowest multipole $\ell=2$ owing to the non-zero $\Lambda$-term, which might be similar to the instability of einsteinian higher dimensional black holes in the de Sitter world \cite{Konoplya:2008au}. Moreover, asymptotically AdS black holes in Einstein-Gauss-Bonnet theory can be unstable even when their radius is large compared to the anti-de Sitter radius $R$, i.e. in the regime of black brane \cite{Konoplya:2017ymp} and detailed analysis of the quasinormal spectrum showed that there is no other instability than the eikonal one \cite{Konoplya2017-preparation}. The instability of ten and eleven dimensional black branes in the Lovelock theory was shown in \cite{Takahashi:2011du}.

Thus, we can see that there is a detailed analysis of the eikonal instability of Einstein-Gauss-Bonnet black holes and branes in $D=5,6$ dimensions mainly and any asymptotic (flat, dS, AdS). At the same time, the results for higher than the second curvature corrections and, thereby, higher $D$ concerns only a few particular cases, such as, e.g., ten and eleven dimensional black branes in AdS \cite{Takahashi:2011du}. Here we shall analyze the general case of the Lovelock gravity allowing for the black hole and brane solutions in various number of spacetime dimensions $D$ and for different asymptotics. The stability analysis is done numerically. We suggest to a reader the Mathematica\textregistered{} code which says whether the black hole metric in any Lovelock theory for given values of all the parameters possesses the eikonal instability or not. It also calculates the line element and the effective potentials numerically with any desired accuracy for both stable and unstable configurations.

The paper is organized as follows. Sec.~\ref{sec:Lovelock} gives detailed description of  the black hole metric in Lovelock theory. Sec.~\ref{sec:perturbations} summarizes the basic information on the master perturbation equations. Sec~\ref{sec:special} is devoted to special cases of fixed coupling constants which allow for the exact solutions to the perturbation equations in terms of hyper-geometrical functions. Sec.~\ref{sec:algorithm} describes the numerical approach to the analysis of the instability regions. In Sec.~\ref{sec:third-order} we determine the parametric region describing the black hole solution with the required asymptotic and limiting regimes at the third-order coupling. Sec.~\ref{sec:results} relates the obtained results on the instability regions and demonstrates several most representative plots in the parametric space. Finally (Sec.~\ref{sec:conclusion}) we summarize the obtained results and discuss possible consequences of the studied instability for higher dimensional gravity and gauge/gravity duality.

\section{Black holes in the Lovelock gravity}\label{sec:Lovelock}

A static spherically symmetric black hole solution to the maximally Gauss-Bonnet extended (Einstein-Lovelock) gravity (\ref{gbg1}) has the form \cite{Wheeler:1985qd}
\begin{equation}\label{Lmetric}
  ds^2=-f(r)dt^2+\frac{1}{f(r)}dr^2 + r^2\,d\Omega_n^2,
\end{equation}
where $d\Omega_n^2$ is a $(n=D-2)$-dimensional sphere, and
\begin{equation}\label{Lfdef}
f(r)=1-r^2\,\psi(r).
\end{equation}
The function $\psi(r)$ satisfies the following relation
\begin{equation}\label{LWdef}
W(\psi(r))\equiv\frac{n}{2}\left(\psi(r)+\sum_{m=2}^\infty\a_m\psi(r)^m\right) - \frac{\Lambda}{n + 1} = \frac{\mu}{r^{n + 1}}\,,
\end{equation}
where
\begin{equation}
\a_m=\frac{\alpha_m}{m}\prod_{p=1}^{2m-2}(n-p)=\frac{\alpha_m}{m}\frac{(n-1)!}{(n-2m+1)!},
\end{equation}
and $\a_m=0$ for any $n\leq2m$ ($m>\m$), implying that $W(\psi)$ is a finite polynomial of $\psi$,
\begin{equation}\label{LWpsi}
W(\psi)=\frac{n}{2}\left(\psi(r)+\sum_{m=2}^{\m}\a_m\psi(r)^m\right) - \frac{\Lambda}{n + 1}.
\end{equation}

Following \cite{Takahashi:2010ye}, we also define a new function $T(r)$ as:
\begin{equation}\label{LTdef}
T(r)\equiv r^{n-1}\frac{dW}{d\psi}=\frac{nr^{n-1}}{2}\left(1+\sum_{m=2}^{\m} m\a_m\psi(r)^{m-1}\right).
\end{equation}

We shall study here the solution to the equation (\ref{LWdef}), which approaches $D$-dimensional Schwarzschild-(anti)-de Sitter in the limit of $\a_m\to0$, that is, we shall consider the branch of solutions which has the Einsteinian limit:
\begin{equation}\label{LEdef}
\psi_0(r)=\frac{2\Lambda}{n(n + 1)}+\frac{2\mu}{nr^{n + 1}}\equiv E(r).
\end{equation}

As one needs to study propagation of spacetime perturbations only outside the black hole, it is useful to fix the system of units in such a way, that the position of the horizon is clear. Therefore, the mass parameter $\mu$ will be measured in units of the black-hole horizon $r_H>0$, so that $f(r_H)=0$. From (\ref{Lfdef}) it follows  that $\psi(r_H)=1/r_H^2$. Then eq. (\ref{LWdef}) yields
\begin{equation}\label{Lmassdef}
  \mu=\frac{n\,r_H^{n-1}}{2}\left(1+\sum_{m=2}^{\m}\frac{\a_m}{r_H^{2m-2}}-\frac{2\Lambda  r_H^2}{n(n+1)}\right).
\end{equation}

\emph{Anti-de Sitter.} In addition, the negative $\Lambda$-term is measured in units of the AdS radius $R$, which is, thereby, defined by
\begin{equation}
\psi(r\to\infty)=-\frac{1}{R^2}.
\end{equation}
Then, from (\ref{LWdef}) one finds
\begin{equation}\label{LAdS}
  \Lambda=-\frac{n(n+1)}{2}\left(\frac{1}{R^2}-\sum_{m=2}^{\m}\frac{(-1)^m\a_m}{R^{2m}}\right).
\end{equation}

\emph{De Sitter.} In the de Sitter case the span of the spatial coordinate $r$ is limited by the cosmological horizon $r_C>r_H$, which we use in order to measure the cosmological constant as
\begin{equation}\label{LdS}
  \Lambda=\frac{n(n+1)}{2}\Biggr(\frac{r_C^{n-1}-r_H^{n-1}}{r_C^{n+1}-r_H^{n+1}}+\sum_{m=2}^{\m}\a_m\frac{r_C^{n-2m+1}-r_H^{n-2m+1}}{r_C^{n+1}-r_H^{n+1}}\Biggr).
\end{equation}
In the limit $r_C\to r_H$ we obtain the extremal value of the cosmological constant, which is given as follows
\begin{eqnarray}\label{LdSe}
  \Lambda_{e}&=&\frac{n(n-1)}{2r_H^2}+\sum_{m=2}^{\m}\a_m\frac{n(n-2m+1)}{2r_H^{2m}}
  \\\nonumber&=&\frac{n(n+1)}{2r_H^2}\left(1+\sum_{m=2}^{\m}\frac{\a_m}{r_H^{2m-2}}\right)-\frac{n}{r_H^2}\left(1+\sum_{m=2}^{\m} m\frac{\a_m}{r_H^{2m-2}}\right).
\end{eqnarray}
Limit $r_C\to\infty$ corresponds to the asymptotically flat spacetime ($\Lambda=0$).

The event horizon $r_H$ satisfies $f'(r_H)>0$, which reads as follows
\begin{equation}\label{Lhorizoncond}
f'(r_H)=-2r_H\psi(r_H)-r_H^2\psi'(r_H)=-\frac{2}{r_H}+\frac{\mu(n+1)}{r_HT(r_H)}=\frac{r_H^n(\Lambda_e-\Lambda)}{T(r_H)} >0,
\end{equation}
where $\Lambda_e$ is given by (\ref{LdSe}). If $T(r_H)>0$ then $f'(r_H)>0$ for $\Lambda<\Lambda_e$ and $r_H>0$.

Since
\begin{equation}\label{monpsi}
  \psi'(r)=-\frac{(n+1)\mu}{r^{n+2}W'(\psi(r))}=-\frac{(n+1)\mu}{r^{3}T(r)},
\end{equation}
we notice that $T(r)$ cannot change the sign outside the event horizon. Indeed, as $T(r)$ is proportional to a polynomial of a finite continuous function $\phi(r)$, if $T(r)$ changes the sign in some point $r_0$, then $T(r_0)=0$ and, due to (\ref{monpsi}), $\psi'(r_0)$ diverges, leading to a naked singularity if $r_0>r_H$. We conclude therefore that outside the event horizon $T(r)>0$ and $\psi(r)$ is monotonically decreasing from $\psi(r_H)=r_H^{-2}$ to either $\psi(r_C)=r_C^{-2}$ (de Sitter) or $\psi(r\rightarrow\infty)=-R^{-2}$ (flat and anti-de Sitter). This implies that the following polynomial is positive definite
\begin{equation}\label{acond}
W'(\psi)\equiv1+\sum_{m=2}^{\m} m\a_m\psi^{m-1}>0,
\end{equation}
if $\psi(r)$ is monotonically decreasing within the following spans:
\begin{equation}\label{psispan}
\begin{array}{rclcc}
r_H^{-2}\geq&\psi&\geq r_C^{-2} &\quad& \mbox{(de Sitter)},\\
r_H^{-2}\geq&\psi&> 0 &\quad&  \mbox{(flat)},\\
r_H^{-2}\geq&\psi&> -R^{-2} &\quad&  \mbox{(anti-de~Sitter)}.
\end{array}
\end{equation}

It is important to notice that the condition (\ref{acond}) allows one to find the physically relevant solution to the equation (\ref{LWdef}). If a solution to (\ref{LWdef}) does not satisfy (\ref{acond}), then we are unable to take limit of $\a_m\to0$ for a fixed radius of the event horizon $r_H$ and chosen asymptotical behavior (i. e., fixed cosmological horizon $r_C$ or AdS radius $R$) without crossing the parametric region in which $W'(\psi)=0$ at some point; the latter leads to a naked singularity. In principle, one can also study the black holes in the parametric region in which there is no einsteinian limit at $\a_m\to0$. However, this nonperturbative configurations are beyond the scope of our study.

Once the above solutions are excluded from consideration, we shall show that in de Sitter space $\Lambda$ must be a monotonic function of the cosmological horizon $r_C\geq r_H$ and, therefore $0\leq\Lambda<\Lambda_e$. Indeed, when $r_C\gg r_H$
$$
  \Lambda\to\frac{n(n+1)}{2}\Biggr(\frac{1}{r_C^2}+\sum_{m=2}^{\m}\frac{\a_m}{r_C^{2m}}\Biggr), \qquad \frac{\partial\Lambda}{\partial r_C}\to-\frac{n(n+1)}{r_C^3}\Biggr(1+\sum_{m=2}^{\m}\frac{\a_m}{r_C^{2m-2}}\Biggr)<0.
$$
Now, suppose that, for some value of $r_H$, $\Lambda$ given by (\ref{LdS}) is not a monotonic function of $r_C$. Then, for the same value of $\Lambda$, there are at least two horizons, ${r_C}_1$ and ${r_C}_2$, such that ${r_C}_2>{r_C}_1>r_H$. This means that the de Sitter space under consideration corresponds to the coordinates $r_H<r<{r_C}_1$ while the next horizon ${r_C}_2$ is another event horizon, i.e. $r_H$ is the inner horizon of a black hole which does not exists in the einsteinian limit.

At the event horizon inequality (\ref{acond}) yields
\begin{equation}\label{Lhorizonbound}
1+\sum_{m=2}^{\m} m\frac{\a_m}{r_H^{2m-2}}>0,
\end{equation}
which can be viewed as a lower bound on the black-hole size $r_H>0$ or, equivalently, the minimal mass of a black hole, which can be calculated using (\ref{Lmassdef}).

From (\ref{LdSe}) it follows that (\ref{Lmassdef}) can be re-written as follows:
\begin{equation}
\mu=r_H^{n-1}\frac{n}{n+1}\left(1+\sum_{m=2}^{\m} m\frac{\a_m}{r_H^{2m-2}}\right)+r_H^{n+1}\frac{\Lambda_e-\Lambda}{n+1},
\end{equation}
so that the mass $\mu$ is positive if $r_H>0$ satisfies the bound (\ref{Lhorizonbound}) and $\Lambda<\Lambda_e$.

If one formally takes the wrong sign of the cosmological constant $\Lambda\geq0$ in the asymptotically AdS space (\ref{LAdS}), then either $\Lambda\geq\Lambda_e$ and $f'(r_H)\leq0$ due to (\ref{Lhorizoncond}), or there is a value of $r_C>r_H$ such that $r_C$ is the cosmological horizon for some other solution $\psi(r)$ of (\ref{LWdef}), which remains asymptotically de Sitter in the limit of $\a_m\to0$, being therefore the relevant branch.

We conclude, therefore, that in the asymptotically AdS case, (\ref{acond}) implies that $\Lambda<0$ and, therefore,
\begin{equation}\label{AdScond}
\sum_{m=2}^{\m}\frac{(-1)^m\a_m}{R^{2m-2}}<1.
\end{equation}
Notice, that substituting the limit $r\to\infty$ ($\psi(r)\to-R^{-2}$) in (\ref{acond}) we find another bound for the parameters $\a_m$ in AdS,
\begin{equation}\label{AdSacond}
\sum_{m=2}^{\m}\frac{(-1)^mm\a_m}{R^{2m-2}}\leq1,
\end{equation}
where the equality in (\ref{AdSacond}) corresponds to asymptotic vanishing of the lefthand side of (\ref{acond}) at the AdS bound.

In Sec.~\ref{sec:third-order} we shall see that in the AdS case inequality (\ref{acond}) imposes a stronger bound on the parameters $\a_m$ than (\ref{AdScond}) and (\ref{AdSacond}).

\section{Perturbation equations}\label{sec:perturbations}
\begin{figure}
\resizebox{\linewidth}{!}{\includegraphics*{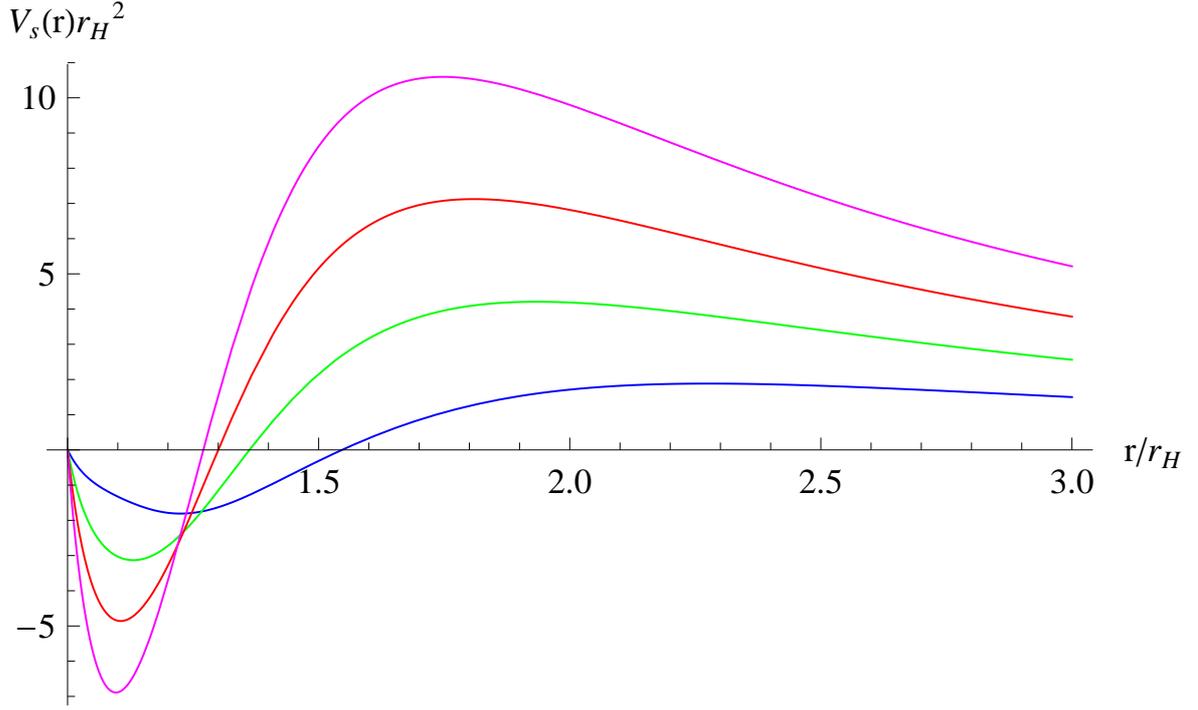}}
\caption{Effective potentials for the scalar-type gravitational perturbations of the Lovelock black hole ($\Lambda =0$, $\a=-0.2 r_{H}^2$, $\b=0.2r_{H}^4$): $\ell=2$ (blue), $\ell=3$ (green), $\ell=4$ (ref), $\ell=5$ (magenta).}\label{fig:pot}
\end{figure}
In \cite{Takahashi:2010ye} it was shown that after the decoupling of angular variables and some algebra, the gravitational perturbation equations can be reduced to the second-order master differential equations
\begin{equation}
\left(\frac{\partial^2}{\partial t^2}-\frac{\partial^2}{\partial r_*^2}+V_i(r_*)\right)\Psi_{i}(t,r_*)=0,
\end{equation}
where $\Psi_i$ are the wave functions, $r_*$ is the tortoise coordinate,
\begin{equation}
dr_*\equiv \frac{dr}{f(r)}=\frac{dr}{1-r^2\psi(r)},
\end{equation}
and $i$ stands for $t$ (\emph{tensor}), $v$ (\emph{vector}), and $s$ (\emph{scalar}) perturbations.
The explicit forms of the effective potentials $V_s(r)$, $V_v(r)$, and $V_t(r)$ are given by
\begin{eqnarray}\nonumber
V_t(r)&=&\frac{\ell(\ell+n-1)f(r)T''(r)}{(n-2)rT'(r)}+\frac{1}{R(r)}\frac{d^2R(r)}{dr_*^2},\\\label{potentials}
V_v(r)&=&\frac{(\ell-1)(\ell+n)f(r)T'(r)}{(n-1)rT(r)}+R(r)\frac{d^2}{dr_*^2}\Biggr(\frac{1}{R(r)}\Biggr),\\\nonumber
V_s(r)&=&\frac{2\ell(\ell+n-1)f(r)P'(r)}{nrP(r)}+\frac{P(r)}{r}\frac{d^2}{dr_*^2}\left(\frac{r}{P(r)}\right),
\end{eqnarray}
where $\ell=2,3,4,\ldots$ is the multipole number, $T(r)$ is given by (\ref{LTdef}), and
$$R(r)=r\sqrt{|T'(r)|},\qquad P(r)=\frac{2(\ell-1)(\ell+n)-nr^3\psi'(r)}{\sqrt{|T'(r)|}}T(r).$$
For large $\ell$ the effective potentials (\ref{potentials}) can be approximated as follows:
\begin{eqnarray}\nonumber
V_t(r)&=&\ell^2\left(\frac{f(r)T''(r)}{(n-2)rT'(r)}+{\cal O}\left(\frac{1}{\ell}\right)\right),\\\label{dominant}
V_v(r)&=&\ell^2\left(\frac{f(r)T'(r)}{(n-1)rT(r)}+{\cal O}\left(\frac{1}{\ell}\right)\right),\\\nonumber
V_s(r)&=&\ell^2\left(\frac{f(r)(2T'(r)^2-T(r)T''(r))}{nrT'(r)T(r)}+{\cal O}\left(\frac{1}{\ell}\right)\right),
\end{eqnarray}
The eikonal instability develops at high multipole numbers $\ell$ and is accompanied by the breakdown of the well-posedness of the initial values problem. In \cite{Takahashi:2010gz} it was shown that once the dominant (at high $\ell$) part of the effective potentials becomes negative, then, the negative gap deepens at higher $\ell$ (see Fig. \ref{fig:pot})  what inevitably leads to the eikonal instability. Therefore, technically our main aim here is to study the parametric regions in which the dominant part of the effective potentials can be negative. This is not a straightforward task, because even the metric function, describing the black hole with required properties and limits cannot be written explicitly in a single analytical expression in general case for the whole range of parameters.

Taking into account that $f(r)\geq0$ and $T(r)>0$, the eikonal instability in the vector channel appears if \cite{Takahashi:2010gz}
\begin{equation}\label{vectorcond}
T'(r)T(r)=r^{2n-3}K(\psi(r))<0,
\end{equation}
If (\ref{vectorcond}) does not hold, eikonal instability can appear either in scalar channel, if
\begin{equation}\label{scalarcond}
M(r)\equiv r^2T(r)^2(2T'(r)^2-T(r)T''(r))=r^{4n-4}J(\psi(r))<0,
\end{equation}
or in tensor channel, if
\begin{equation}\label{tensorcond}
N(r)\equiv r^2T(r)^3T''(r)=r^{4n-4}L(\psi(r))<0.
\end{equation}
Here, following \cite{Takahashi:2010gz} we defined\footnote{Note, in \cite{Takahashi:2010gz} $J(\psi)$ is denoted as $M(\psi)$, and $W(\psi)$ in (\ref{LWpsi}) differs by the positive factor $n/2$.} the above three functions as follows:
\begin{eqnarray}
J(\psi)&\equiv& n(n-1)W'(\psi)^4- 3n(n+1)W(\psi)W'(\psi)^2W''(\psi)\nonumber\\
&&+(n+1)^2W(\psi)^2\left(3W''(\psi)^2-W'(\psi)W'''(\psi)\right),\nonumber\\
\label{XYZdef}
K(\psi)&\equiv& (n-1)W'(\psi)^2-(n+1)W(\psi)W''(\psi),\\
L(\psi)&\equiv& (n-1)(n-2)W'(\psi)^4-(n+1)(n-4)W(\psi)W'(\psi)^2W''(\psi)+ \nonumber\\
&&+(n+1)^2W(\psi)^2\left(W'(\psi)W'''(\psi)-W''(\psi)^2\right).\nonumber
\end{eqnarray}
Unlike Gauss-Bonnet case \cite{Konoplya:2017ymp}, these functions are not polynomials of $r$, with a minimum at $r=r_H$. However, the problem of instability is reduced now to finding the parametric region for which one of the polynomials, $J(\psi)$, $K(\psi)$, or $L(\psi)$, is negative for some values of $\psi$ in (\ref{psispan}).

The parametric region for which $K(\psi)$ is negative corresponds to the so-called \emph{ghost instability}: the kinetic term of perturbations has a wrong sign in this region \cite{Takahashi:2010gz}. If $K(\psi)$ is positive and either $J(\psi)$ or $L(\psi)$ is negative, then one has dynamical instability in the scalar ($J(\psi)<0$) or tensor ($L(\psi)<0$) channel, respectively \cite{Takahashi:2010gz}.
Notice, that $J(\psi)$ and $L(\psi)$ cannot be negative at the same point, because
\begin{equation}
J(\psi)+L(\psi)=2K(\psi)^2\geq0.
\end{equation}
In addition, the following polynomials are positive in the parametric region under consideration
$$W(\psi)=W(\psi(r))=\frac{\mu}{r^{n+1}}>0,\qquad W'(\psi)=W'(\psi(r))=\frac{T(r)}{r^{n-1}}>0.$$

\section{Special case $T(r) = const$.}\label{sec:special}
When $T(r) = const$, the above argumentation fails. Therefore, here we consider separately a special case of dimensionally continued BTZ black holes \cite{Banados:1993ur} in odd dimensions, $n=2k-1$, with the coupling constants, fixed as follows \cite{Gannouji:2013eka}
\begin{equation}
\a_m=\frac{(k-1)!}{m!(k-m)!}R^{2m-2}.
\end{equation}
Then, one has
\begin{equation}
W(\psi)=-\frac{\Lambda}{2k}\left(1-\frac{2k-1}{\Lambda}\psi\right)^k=\frac{2k-1}{2k R^2}\left(1+R^2\psi\right)^k.%\left(\a_m=\binom{k}{m}\frac{R^{2m-2}}{k}\right),
\end{equation}
From (\ref{LWdef}) we find that
\begin{equation}\label{psiex}
1+R^2\psi(r)=\frac{1}{r^2}\left(\frac{2k R^2\mu}{2k-1}\right)^{1/k}=\frac{r_H^2+R^2}{r^2},
\end{equation}
and, thereby, the function
\begin{equation}\label{Tconst}
T(r)=r^{2k-2}\frac{2k-1}{2}\left(1+R^2\psi(r)\right)^{k-1}=\left(\frac{k}{2}-1\right)\left(\frac{2k R^2\mu}{2k-1}\right)^{1-\frac{1}{k}}
\end{equation}
is a constant, leading to indeterminate expressions for tensor and scalar potentials in (\ref{dominant}). Indeed, in this case $T'(r)=0$ and $T''(r)=0$, however, their quotient remains finite. In order to see this we differentiate (\ref{Tconst}), substitute (\ref{monpsi}), and, taking into account (\ref{psiex}), obtain
\begin{equation}\label{Tind}
\frac{T''(r)}{T'(r)}=\frac{2k-3}{r}+\frac{4R^2}{2k-1}\frac{\mu}{r^{2k+1}\left(1+R^2\psi(r)\right)^k}=\frac{2k-1}{r}=\frac{n}{r}.
\end{equation}
Substituting (\ref{Tind}) into (\ref{potentials}), we find
\begin{eqnarray}\nonumber
V_t(r)&=&f(r)\left(\frac{n(n+2)}{4r^2}f(r)+\frac{n+2}{2r}f'(r)+\frac{\ell(\ell+n-1)n}{(n-2)r^2}\right),\\\label{potTconst}
V_v(r)&=&f(r)\left(\frac{(n+2)(n+4)}{4r^2}f(r)-\frac{n+2}{2r}f'(r)\right),\\\nonumber
V_s(r)&=&f(r)\left(\frac{n(n+2)}{4r^2}f(r)+\frac{n+2}{2r}f'(r)-\frac{\ell(\ell+n-1)}{r^2}\right),
\end{eqnarray}
where $f(r)=1-r^2\psi(r)=(r^2-r_H^2)/R^2$.

The case $T(r) = const$ is remarkable also from a different point of view.
Usually it is not easy to find an exact solution of the perturbation equations, though several exceptions exist. %\cite{Staicova:2014ioa,Fiziev:2011mm,exactQNM}.
In \cite{Gonzalez:2017gwa} in a similar fashion with \cite{Gonzalez:2010vv} exact solutions for the perturbation equations in terms of hypergeometric functions were obtained for the case of five-dimensional Gauss-Bonnet-AdS black holes with the fixed coupling constant $\alpha_2=R^2/2$.
We believe that the eigenvalue problem in this more general case  of the dimensionally continued BTZ black holes can be solved analytically in a similar manner as for $n=3$ in \cite{Gonzalez:2010vv}. For us, however, it is important only that the dominant (at high $\ell$) part of the effective potentials (\ref{potTconst}) say us that there must be an evident eikonal instability in the scalar sector while the perturbations of vector type and tensor type are linearly stable.

\section{Numerical analysis of instability regions in the most general case}\label{sec:algorithm}
Now we are in a position to describe the algorithm for the numerical analysis of (the relevant branch of) the Lovelock black hole.
\begin{enumerate}
\item Accroding to (\ref{psispan}), for given values of $r_H$ and $r_C$ (de Sitter) or $R$ (AdS) the interval for the values of $\psi$ are:  $r_H^{-2}\geq\psi\geq r_C^{-2}$ (de Sitter and, at $r_C \rightarrow \infty$, flat cases) or $r_H^{-2}\geq\psi> -R^{-2}$ (anti-de Sitter).
\item For given values of $\a_2,\a_3,\a_4,\ldots$ from (\ref{Lmassdef}) we determine the mass parameter $\mu$, while from (\ref{LdS}) (for de Sitter) or (\ref{LAdS}) (for anti-de Sitter) we find the corresponding value of the $\Lambda$-term.
\item Using the obtained value of the $\Lambda$-term, from (\ref{LWdef}) we find the polynomial $W(\psi)$ and check whether $W'(\psi)>0$ in the interval of changing of $\psi$ (\ref{psispan}). Otherwise, we are outside the parametric region, describing a black hole with the required properties.
\item Using (\ref{XYZdef}), we calculate $K(\psi)$ and check if it is zero. If so, we have a special case, considered in Sec.~\ref{sec:special}, for which there is the eikonal instability in the scalar sector.
\item We check if $K(\psi)\geq0$ in the interval (\ref{psispan}). Otherwise, one has the ghost instability.
\item From (\ref{XYZdef}) we find $J(\psi)$ and $L(\psi)$. If there is a point for which $J(\psi)$ or $L(\psi)$ are negative, we have the eikonal instability in the scalar or tensor channel, respectively.
\item If $W'(\psi)>0$ we tabulate the interval for $\psi$ and, using (\ref{LWdef}), calculate corresponding values of $r$. We find an approximate function $\psi(r)$ using spline interpolation of the resulting table.
\item In order to find accurate function $\psi(r)$ we solve numerically (\ref{LWdef}) with the initial guess given by the spline interpolation. In this way we can find the value of $\psi$ with arbitrary precision.
\item Once $\psi(r)$ is known we can calculate $T(r)$ using (\ref{LTdef}), and, finally, the effective potentials (\ref{potentials}).
\end{enumerate}

The presence of a negative gap, growing with $\ell$, so that the potentials in the eikonal regime becomes negatively dominant, indicates the instability. This way, for any given values of the incoming parameters, such as, the black hole radius $r_{H}$, radius of the cosmological horizon $r_{C}$ (or anti-de Sitter radius $R$), the number of spacetime dimensions $D=n+2$, coupling constants $\a_2,\a_3,\a_4,\ldots$, one can answer the two questions:
\begin{itemize}
\item whether the metric describes a black hole and has the required Einsteinian limit, and
\item if it does, then, whether such a black hole has eikonal instability or not.
\end{itemize}
This numerical method is implemented in Wolfram~\emph{Mathematica}\textregistered{} (supplementary file).

\section{Black-hole's parametric region for the third order Lovelock theory}\label{sec:third-order}

In this section we shall consider the case of the third order Einstein-Lovelock theory, for which considerable part of the analysis can be done analytically. At the third order Lovelock theory, that is, once $\a_m=0$ for $m>3$ we can rewrite eq. (\ref{LWdef}) as follows:
\begin{equation}\label{Wdef}
W[\psi(r)]\equiv\frac{n}{2}\left(\b\psi(r)^3+\a\psi(r)^2+\psi(r)\right) - \frac{\Lambda}{n + 1} = \frac{\mu}{r^{n + 1}},
\end{equation}
where
$$\a \equiv \a_2 \equiv \alpha_2\frac{(n-1)(n-2)}{2}, \qquad \b \equiv \a_3 \equiv \alpha_3\frac{(n-1)(n-2)(n-3)(n-4)}{3}.$$

Then, the bound for the black-hole size (\ref{horizonbound}) reads
\begin{equation}\label{horizonbound}
1+\frac{2\a}{r_H^2}+\frac{3\b}{r_H^4}>0,
\end{equation}
implying that
\begin{equation}\label{horizonboundexplicit}
r_H>\left\{
      \begin{array}{lll}
        0, & \a\geq0, & \b\geq0; \\
        \sqrt{\sqrt{\a^2-3\b}-\a}, & \a\geq0, &\b<0; \\
        0, & \a<0, & \b>\dfrac{\a^2}{3}; \\
        \sqrt{-\a\left(1+\sqrt{1-\dfrac{3\b}{\a^2}}\right)}, & \a<0, & \b\leq\dfrac{\a^2}{3}.
      \end{array}
    \right.
\end{equation}

\begin{figure}
\resizebox{\linewidth}{!}{\includegraphics*{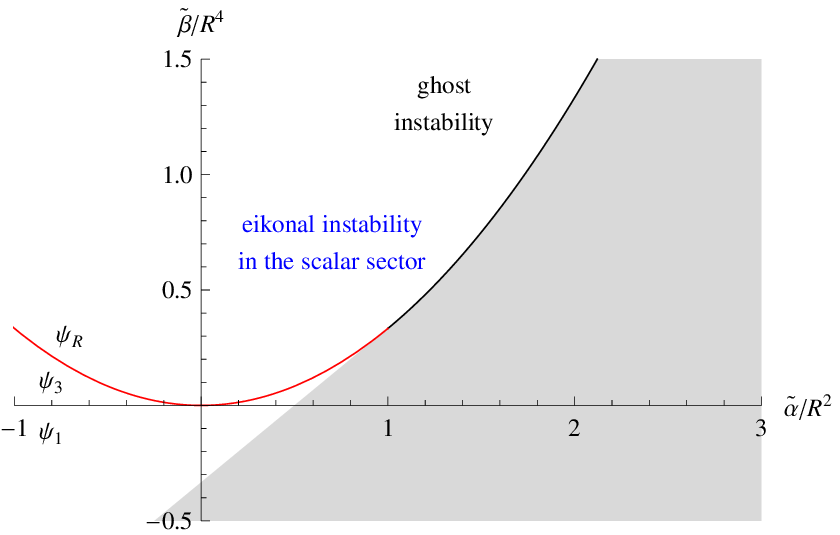}}
\caption{Explicit form of the relevant solution in the parametric region of the AdS black holes. The lower right region is excluded due to (\ref{alphaboundexplicit}). The red parabola $3\b=\a^2\leq R^4$ corresponds to the solution (\ref{alphalimit}). For $n=5$ the right limit of the red line ($3\b=R^4$, $\a=R^2$) corresponds to the special solution considered in Sec.~\ref{sec:special}. The black line $3\b=\a^2>R^4$ is excluded due to (\ref{alphaboundexplicit}). Above the parabola the solution is given by (\ref{psi}), below the parabola we choose $\psi_3$ for $\b>0$ and $\psi_2$ for $\b<0$ of the Tschirnhaus-Vieta form (\ref{psis}).}\label{fig:abAdS}
\end{figure}

The condition (\ref{horizonboundexplicit}) is sufficient to provide the positiveness of the polynomial (\ref{acond}) in the flat and de Sitter case. In the AdS case when $\b\leq\a^2/3$ we have an additional constrain, which can be found from the condition that the polynomial  (\ref{acond}) is positive also for negative values of $\psi>-R^{-2}$,
\begin{equation}\label{alphaboundexplicit}
\a+\sqrt{\a^2-3\b}\leq R^2.
\end{equation}
The condition (\ref{alphaboundexplicit}) is stronger than (\ref{AdScond}) and (\ref{AdSacond}). The unstable parametric region is presented on Fig.~\ref{fig:abAdS}.

In order to write down the relevant solution representing the required black-hole metric, we introduce the new function
\begin{equation}\label{Fdef}
F(r)=\frac{27\b^2}{2\a^3} E(r)+\frac{9\b}{2\a^2}-1,
\end{equation}
where $E(r)$ is given by (\ref{LEdef})
$$E(r)=\frac{2\Lambda}{n(n + 1)}+\frac{2\mu}{nr^{n + 1}},$$
and the mass parameter is defined by (\ref{massdef}) as follows
\begin{equation}\label{massdef}
  \mu=\frac{n\,r_H^{n-1}}{2}\left(1+\frac{\a}{r_H^2}+\frac{\b}{r_H^4}-\frac{2\Lambda  r_H^2}{n(n+1)}\right)>0.
\end{equation}

For $\b\geq\a^2/3$ there is only one real solution of (\ref{Wdef})
\begin{equation}\label{psi}
\psi_R(r)=\frac{\a}{3\b}\left(\sqrt[3]{\sqrt{F(r)^2+\left(\frac{3\b}{\a^2}-1\right)^3}+F(r)}-\sqrt[3]{\sqrt{F(r)^2+\left(\frac{3\b}{\a^2}-1\right)^3}-F(r)}-1\right).
\end{equation}
In particular, if $\b=\a^2/3$, the solution reads
\begin{equation}\label{alphalimit}
\psi_R(r)=\frac{1}{\a}\left(\sqrt[3]{1+3\a E(r)}-1\right)=\frac{3E(r)}{1+\sqrt[3]{1+3\a E(r)}+\sqrt[3]{(1+3\a E(r))^2}}.
\end{equation}
In the limit $\a\to0$ ($\b>0$) we find
\begin{equation}\label{betalimit}
\psi_R(r)=\sqrt[3]{\sqrt{\left(\frac{E(r)}{2\b}\right)^2+\frac{1}{27\b^3}}+\frac{E(r)}{2\b}}-\sqrt[3]{\sqrt{\left(\frac{E(r)}{2\b}\right)^2+\frac{1}{27\b^3}}-\frac{E(r)}{2\b}}.
\end{equation}

For $\b<\a^2/3$ there are three real solutions to the cubic equation (\ref{Wdef}), which can be found using the Tschirnhaus-Vieta approach
\begin{equation}\label{psis}
\psi_n(r)=\frac{2\a}{3\b}\sqrt{1-\frac{3\b}{\a^2}}\cos\Biggr(\frac{1}{3}\arccos\left(\frac{F(r)}{(1-3\b/\a^2)^{3/2}}\right)+\frac{2\pi n}{3}\Biggr)-\frac{\a}{3\b}.
\end{equation}
It is important to notice that, in order to obtain the appropriate black-hole metric, we choose different roots of (\ref{psis}). In the limit $\b\to\a^2/3$ the solution (\ref{alphalimit}) is reproduced by $\psi_3(r)$ for $F(r)>0$ and $\psi_1(r)$ for $F(r)<0$.

Substituting (\ref{massdef}) into (\ref{Fdef}), in the limit $\b\to\a^2/3$ we find
$$F(r)=\frac{1}{2}+\frac{\a^3}{2r_H^6}\left(\frac{r_H}{r}\right)^{n+1}+\frac{3\a}{2r_H^2}\left(\frac{r_H}{r}\right)^{n+1}+\frac{3\a^2}{2r_H^4}\left(\frac{r_H}{r}\right)^{n+1}+\frac{3\a\Lambda}{n(n+1)}-\frac{3\a\Lambda}{n(n+1)}\left(\frac{r_H}{r}\right)^{n+1},$$
which is a monotonic function, because its derivative does not change the sign at a fixed $\a$:
$$F'(r)=-\frac{3(n+1)\a r_H^{n-1}}{2r^{n+2}}\left(1+\frac{\a}{r_H^2}+\frac{\a^2}{3r_H^4}-\frac{2\Lambda r_H^2}{n(n+1)}\right)=-\frac{3(n+1)\a\mu}{nr^{n+2}}.$$
If $\a\leq0$ the function $F(r)$ is nondecreasing. Therefore, since (\ref{horizonboundexplicit}) implies that $r_H^2>-\a$, then we have
$$F(r)\geq F(r_H)=\frac{1}{2}\left(1+\frac{\a}{r_H^2}\right)^3>0.$$
If $\a>0$, when $\Lambda\geq0$ the function $F(r)$ is positive for any $r\geq r_H$, and, when $\Lambda<0$, one has
\begin{equation}\label{lambdaa}
F(r)>\lim_{r\to\infty}F(r)=\frac{1}{2}+\frac{3\a\Lambda}{n(n+1)}=\frac{1}{2}\left(1-\frac{\a}{R^2}\right)^3.
\end{equation}
Thus, for $\b=\a^2/3$ in the asymptotically de Sitter, flat, and anti-de Sitter (if $\a\leq R^2$) backgrounds we have
\begin{equation}\label{Feq}
F(r)=\frac{1+3\a E(r)}{2}>0.
\end{equation}
Since the region $\a>R^2$ is excluded due to (\ref{alphaboundexplicit}), we conclude that the relevant solution for $\b\to\a^2/3$ is $\psi_3(r)$.

In the limit $\b\to0$ we obtain the correct Gauss-Bonnet solution for
$$\lim_{\b\to-0}\psi_2=\lim_{\b\to+0}\psi_3=\frac{1}{2\a}\left(\sqrt{1+4\a E(r)}-1\right)=\frac{2E(r)}{1+\sqrt{1+4\a E(r)}}=\psi_{GB}(r).$$
The other one-sided limits correspond to wrong signs of the square root. We conclude therefore that the relevant black-hole solution is given by (see Fig.~\ref{fig:abAdS} for AdS black holes)
\begin{equation}\label{psisb}
\psi(r)=\left\{
    \begin{array}{ll}
      \psi_R(r), & \b\geq\a^2/3; \\
      \psi_3(r), & 0<\b<\a^2/3; \\
      \psi_{GB}(r), & \b=0; \\
      \psi_2(r), & \b<0.
    \end{array}
  \right.
\end{equation}

Unfortunately, this kind of analytical analysis becomes too much involved when considering the higher than the third order terms, so that in the general case we have to use the above discussed numerical treatment.

\section{The portrait of eikonal instability}\label{sec:results}

Let us start from the description of the ghost instability in the third order Lovelock theory. First, we observe that for $\b\leq\a^2/3$ (\ref{horizonboundexplicit}) and (\ref{alphaboundexplicit}) in AdS yield
\begin{equation}\label{alphapsi}
  1+\a\psi(r)>0.
\end{equation}
Indeed, $\a\leq0$, $\psi(r)\leq r_H^{-2}<-\a^{-1}$ due to (\ref{horizonboundexplicit}); if $\a>0$, then $\psi(r)>-R^{-2}\geq-\a^{-1}$ due to (\ref{alphaboundexplicit}).

Now we are in a position to analyze the ghost instability region, i.~e., the parametric region in which $K(\psi)$ can be negative. From (\ref{XYZdef}) we notice that if $K(\psi)<0$, then
\begin{equation}\label{Wpp}
  W''(\psi)=n(\a+3\b\psi)>\frac{n-1}{n+1}\frac{W'(\psi)^2}{W(\psi)}>0.
\end{equation}

Therefore, for $\b\leq\a^2/3$
\begin{eqnarray}\label{Keq}
 K(\psi)&=&\frac{n^2(n-5)}{12}\left(1+2\a\psi+3\b\psi^2\right)^2+\\\nonumber
&&+\frac{n^2(n+1)}{6}\left(1+\a\psi+\frac{6\Lambda(\a+3\b\psi)}{n(n+1)}+(\a^2-3\b)\psi^2\right),
%K(\psi)=\frac{n}{2}\left(\frac{n(n-1)}{2} + 2 \a \Lambda + (\a n(n-3) + 6 \b \Lambda) \psi + (\a^2(n-3)-6 \b) n \psi^2 + 2\a\b n(n-5) \psi^3 +  \b^2 \frac{3n(n-5)}{2} \psi^4\right).
\end{eqnarray}
is positive, if $\Lambda\geq0$ due to (\ref{alphapsi}) and (\ref{Wpp}).

If $\Lambda<0$, using (\ref{LAdS}) we find that the following expression
\begin{eqnarray}
&&1+\a\psi+\frac{6\Lambda(\a+3\b\psi)}{n(n+1)}+(\a^2-3\b)\psi^2=\\\nonumber
&&\qquad-3\a\frac{\b-\a R^2+R^4}{R^6} + 1 + \left(\a-9\b\frac{\b - \a R^2 + R^4}{R^6}\right) \psi + (\a^2 - 3\b)\psi^2
\end{eqnarray}
is nonnegative for any $\psi$ if
$$2\a-R^2\leq\frac{3\b}{R^2}\leq\a+2\left(\sqrt{\a^2-\a R^2+R^4}-R^2\right).$$
When
$$\frac{3\b}{R^2}>\a+2\left(\sqrt{\a^2-\a R^2+R^4}-R^2\right),$$
the quadratic form can be negative if
\begin{eqnarray}\nonumber
\psi&>&\frac{9 \b^2 - \a R^6 - 9 \b R^2 (\a - R^2)-(3 \b - 2 \a R^2 + R^4)\sqrt{3 (3 \b^2 - 2 \a \b R^2 - \a^2 R^4 + 4 \b R^4)}}{2 (\a^2 - 3 \b) R^6}\\\nonumber
&>&\frac{1}{\sqrt{\a^2-3\b}-\a},
\end{eqnarray}
what is never satisfied since (\ref{horizonboundexplicit}) implies that
$$\psi(r)\leq\psi(r_H)=\frac{1}{r_H^2}<\frac{1}{\sqrt{\a^2-3\b}-\a}.$$

\begin{figure}
\resizebox{\linewidth}{!}{\includegraphics*{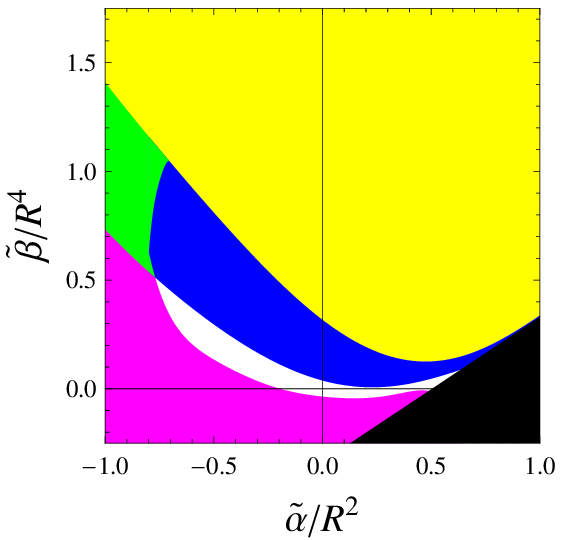}\includegraphics*{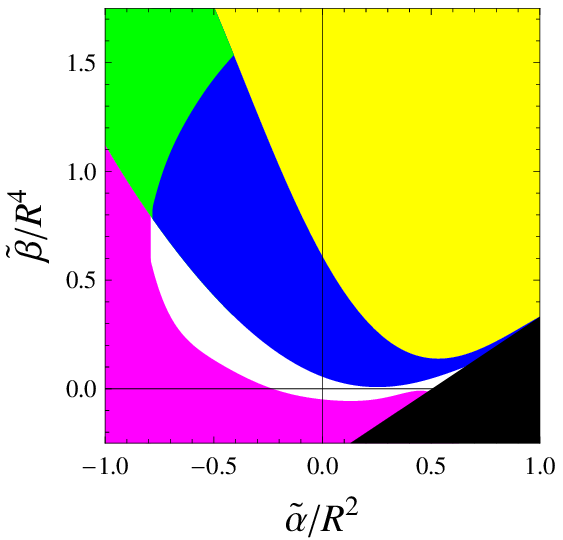}\includegraphics*{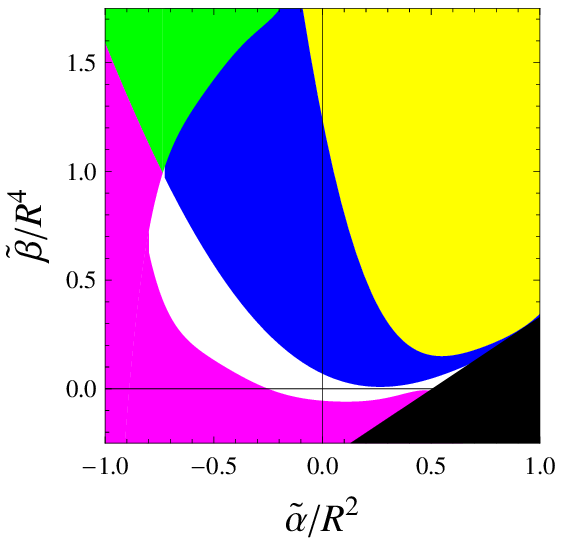}\includegraphics*{n5brane.eps}}
\caption{Instabilities for the AdS black holes ($n=5$) for $r_H=2R$, $r_H=3R$, $r_H=5R$, $r_H=\infty$ (from left to right). Black region is the excluded parametric region, yellow - ghost instability, blue - eikonal instability in the scalar channel, magenta - eikonal instability in the tensor channel, green - eikonal instability in both tensor and scalar channel.}\label{fig:AdS}
\end{figure}

Therefore, we conclude that the ghost instability appears only for $\b>\a^2/3$. It is interesting to note that for $\b>\a^2/3$ inequality (\ref{horizonbound}) is satisfied for any $r_H>0$. At the same time, the parametric region of ghost instability expands as $r_H$ decreases (see Fig.~\ref{fig:AdS}), so that too small black holes do not exist for $\b>\a^2/3$ as well.

The case $n=5$ corresponds to the largest region of the ghost instability. Indeed, for $n<5$ there is no ghost instability because $\b=0$. For $n>5$ the first term in (\ref{Keq}) is positive, implying that the ghost instability region shrinks as $n$ increases.

Once there is no ghost instability, the eikonal instability appears in the scalar channel if and only if $J(\psi)<0$ for some point outside the event horizon.
From (\ref{XYZdef}) we notice that $J(\psi)$ is a quadratic polynomial of $W(\psi)$, being nonnegative provided
\begin{eqnarray}\label{scalar1}
    &&n(n-4)(\a^2-3\b)+2\b(n-10)W'(\psi)\geq0,\\\label{scalar2}
    &&2\a^2-\b+10\a\b\psi+15\b^2\psi^2\geq0.
\end{eqnarray}
Taking into account that $W'(\psi)>0$ and (\ref{scalar2}) is satisfied for any $\b\leq\a^2/3$, we observe that there is no eikonal instability for $\b\leq0$ and $4\leq n<10$. For $n\geq10$ we use the explicit form of (\ref{scalar1}) and notice that the quadratic form is nonnegative for any $\psi$ if $\b\leq\a^2/3$.

Thus, we conclude that the eikonal instability in the scalar sector exists only for
\begin{equation}\b>\left\{
       \begin{array}{ll}
         0, & 5\leq n<10; \\
         \dfrac{\a^2}{3}, & n\geq10.
       \end{array}
     \right.
\end{equation}

From the analysis of the regions of instability we see that if we start from any point in the stable parametric region and increase $\b$ we first go to the region of the eikonal instability in the scalar sector and, for sufficiently large value of $\b$, we come to the region of ghost instability. In other words, the region of the ghost instability is always contained inside the region of the eikonal instability.

\begin{figure}
\resizebox{\linewidth}{!}{\includegraphics*{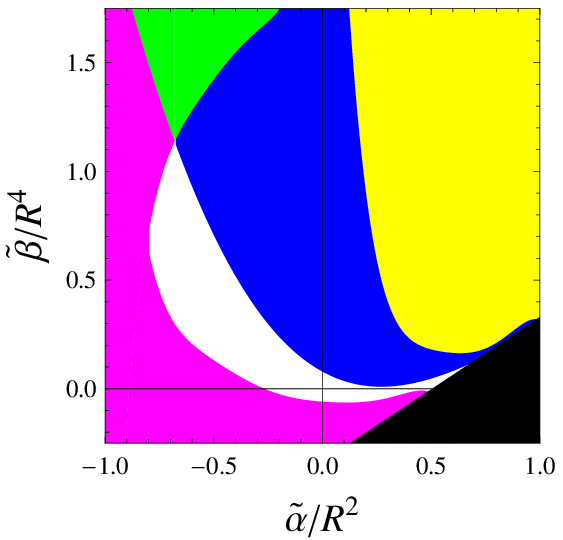}\includegraphics*{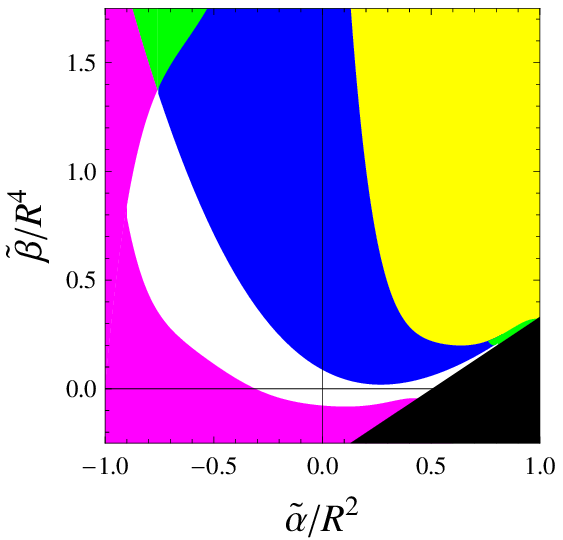}\includegraphics*{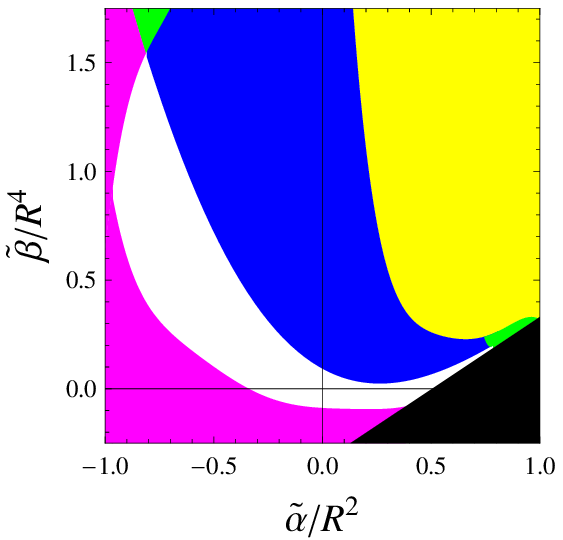}\includegraphics*{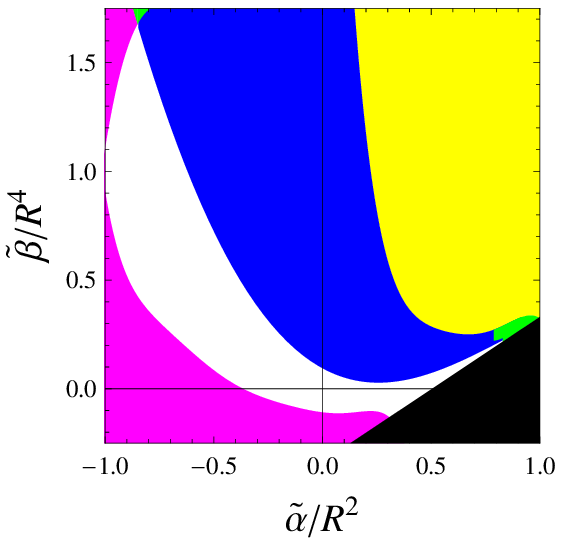}}
\caption{Instabilities for the AdS black brane ($r_H\to\infty$) for $n=5$, $n=6$, $n=7$, $n=8$ (from left to right). Black region is the excluded parametric region, yellow - ghost instability, blue - eikonal instability in the scalar channel, magenta - eikonal instability in the tensor channel, green - eikonal instability in both tensor and scalar channel.}\label{fig:branes}
\end{figure}

The region of instability of the $D=7, 8$-dimensional AdS black branes can be fully depicted on the two dimensional plots (see fig.~\ref{fig:branes}), where we also show $D=9, 10$ cases at $\a_4 =0$. Instability regions of the asymptotically AdS $D=8$-dimensional black hole at various values of the black hole radius $r_H$ is shown on fig.~\ref{fig:AdS}. There one can see that when the black hole becomes smaller, the region of stability shrinks, mainly owing to the increasing region of the instability in the scalar channel. The region of the ghost instability also increases when $r_H$ is decreasing, but still stays within the eikonal instability region. The region of instability in the tensor channel is deformed relatively softly for small black holes.

\begin{figure}
\resizebox{\linewidth}{!}{\includegraphics*{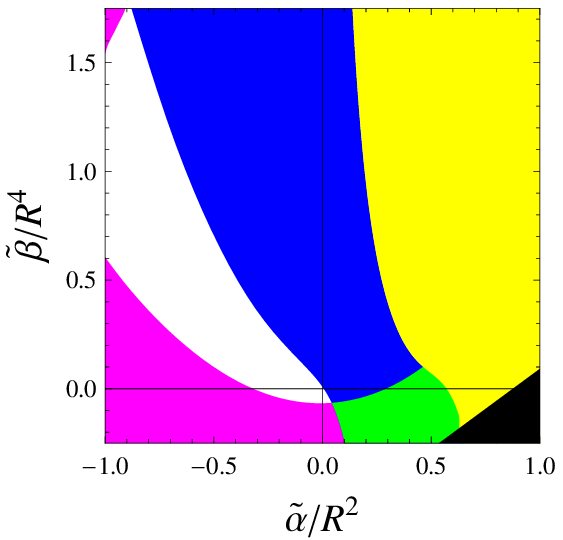}\includegraphics*{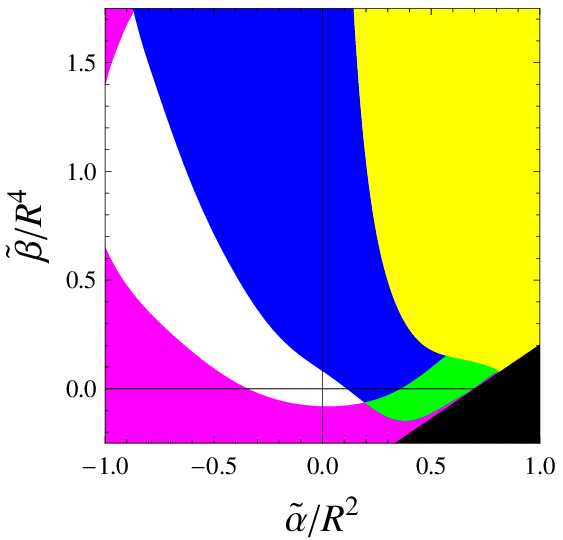}\includegraphics*{n8brane.eps}\includegraphics*{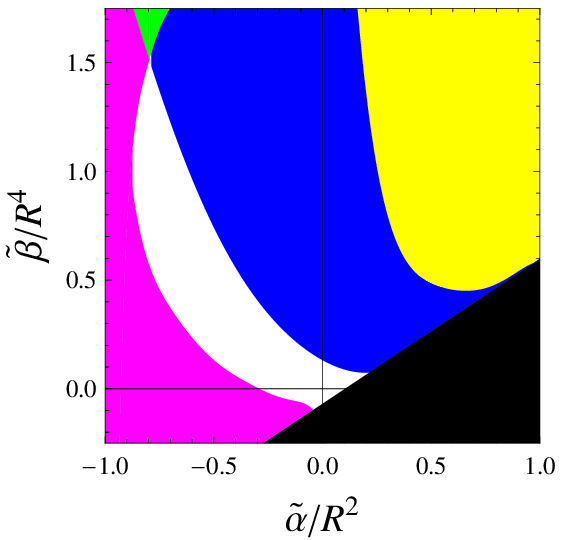}}
\caption{Instabilities for the AdS black brane ($r_H\to\infty$) for $n=8$, $\a_4=-0.2R^6$, $\a_4=-0.1R^6$, $\a_4=0$, $\a_4=0.2R^6$ (from left to right). Black region is the excluded parametric region, yellow - ghost instability, blue - eikonal instability in the scalar channel, magenta - eikonal instability in the tensor channel, green - eikonal instability in both tensor and scalar channel.}\label{fig:branes4}
\end{figure}

\begin{figure}
\resizebox{\linewidth}{!}{\includegraphics*{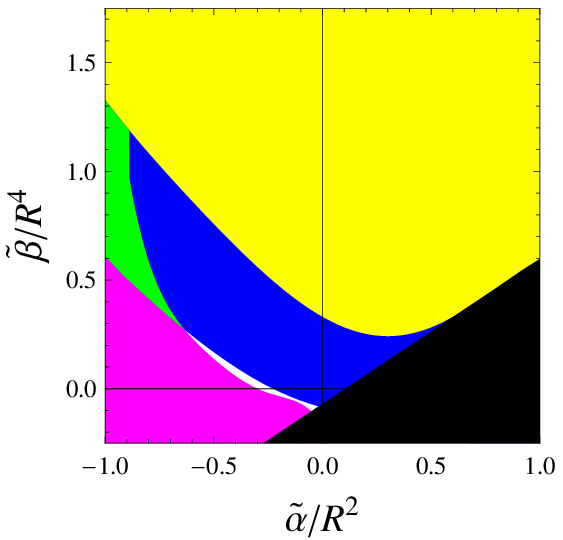}\includegraphics*{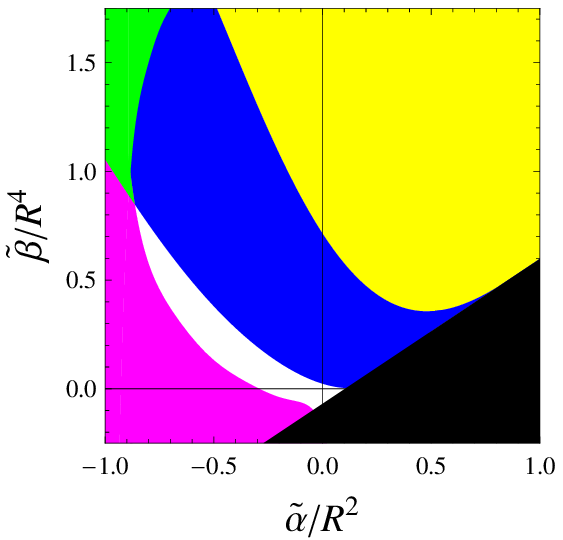}\includegraphics*{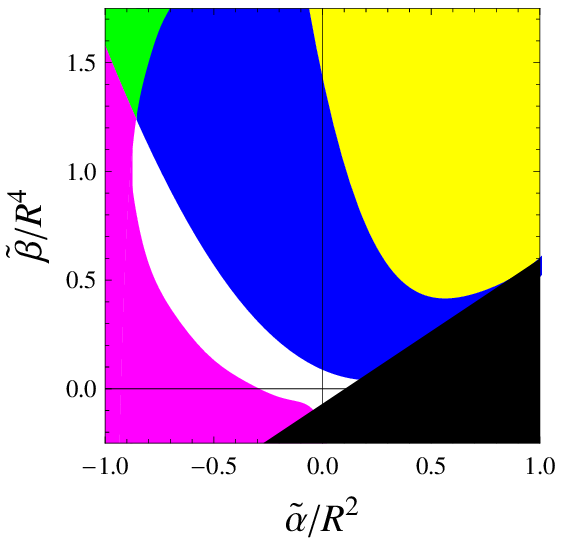}\includegraphics*{n8brane+0_2.eps}}
\caption{Instabilities for the AdS black holes ($n=8$, $\a_4=0.2R^6$) for $r_H=2R$, $r_H=3R$, $r_H=5R$, $r_H=\infty$ (from left to right). Black region is the excluded parametric region, yellow - ghost instability, blue - eikonal instability in the scalar channel, magenta - eikonal instability in the tensor channel, green - eikonal instability in both tensor and scalar channel.}\label{fig:AdS4}
\end{figure}

On fig.~\ref{fig:branes4} one can see how the region of instability at $\a_4 =0$ of the $D=10$-dimensional asymptotically AdS black brane is deformed by both positive and negative values of $\a_4$. Notice, that the parametric region, describing a black hole with required properties, shrinks considerably when increasing  $\a_4$. On fig.~\ref{fig:AdS4} one can see the region of instability for $D=10$ AdS black hole at various values of the radius $r_H$. Here, again the stability region dramatically shrinks when going over to smaller black holes.

\begin{figure}
\resizebox{\linewidth}{!}{\includegraphics*{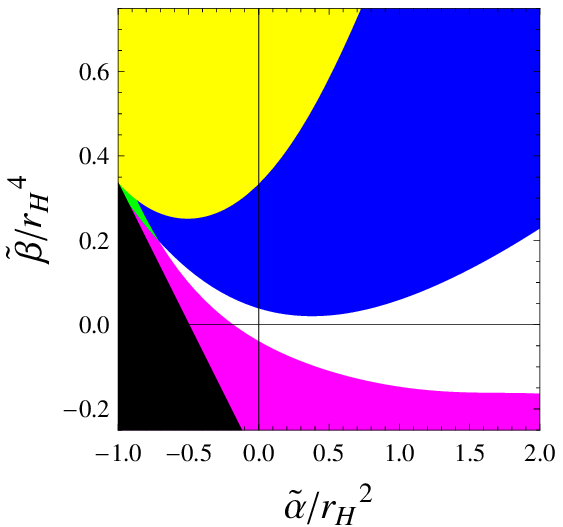}\includegraphics*{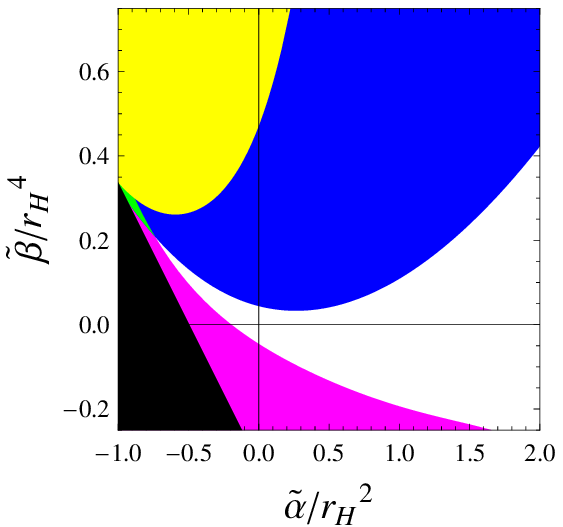}\includegraphics*{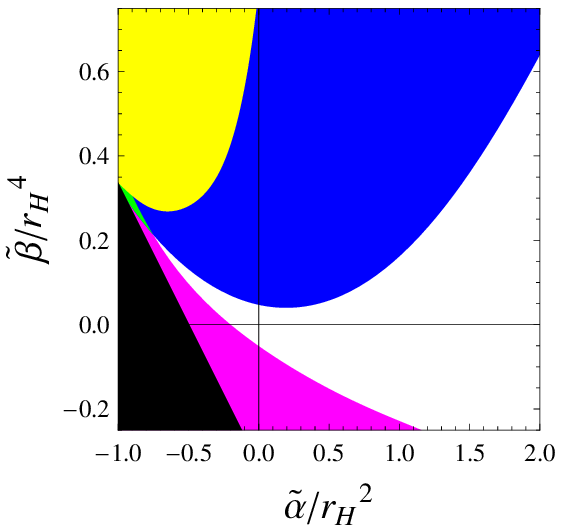}\includegraphics*{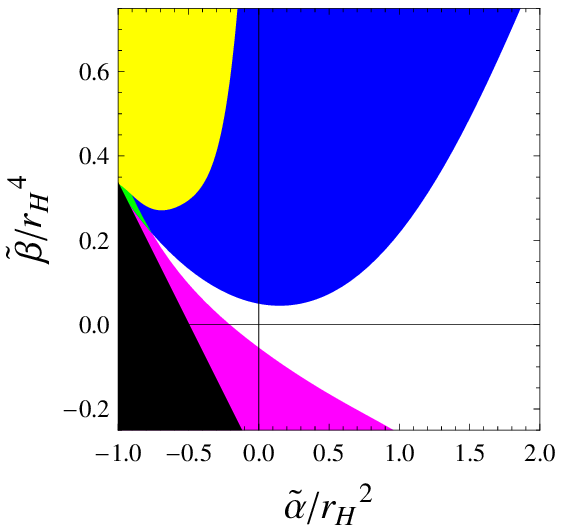}}
\caption{Instabilities for black holes in the flat space ($\Lambda=0$) for $n=5$, $n=6$, $n=7$, $n=8$ (from left to right). Black region is the excluded parametric region, yellow - ghost instability, blue - eikonal instability in the scalar channel, magenta - eikonal instability in the tensor channel, green - eikonal instability in both tensor and scalar channel.}\label{fig:flat}
\end{figure}

\begin{figure}
\resizebox{\linewidth}{!}{\includegraphics*{n5flat.eps}\includegraphics*{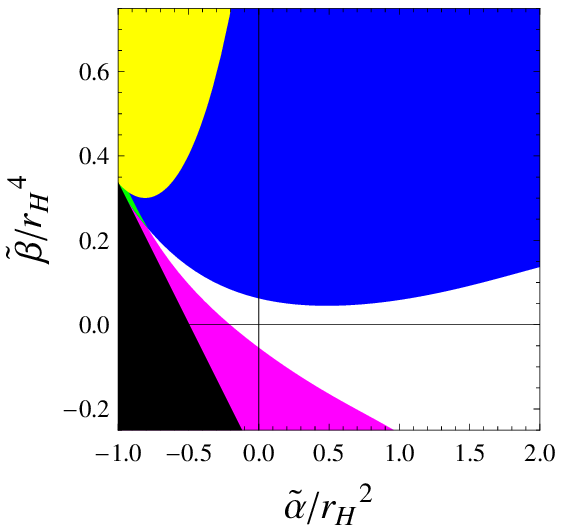}\includegraphics*{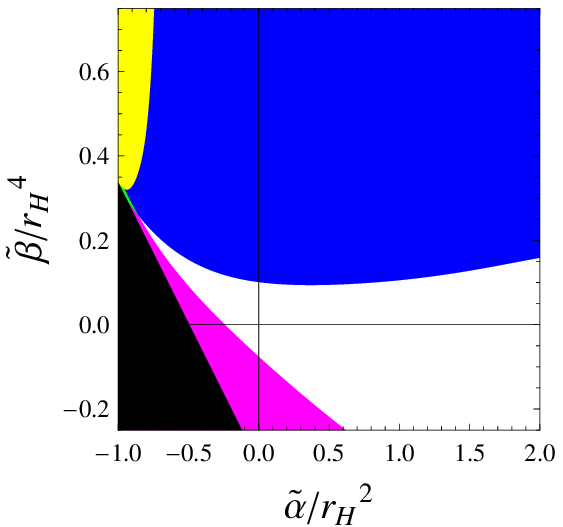}\includegraphics*{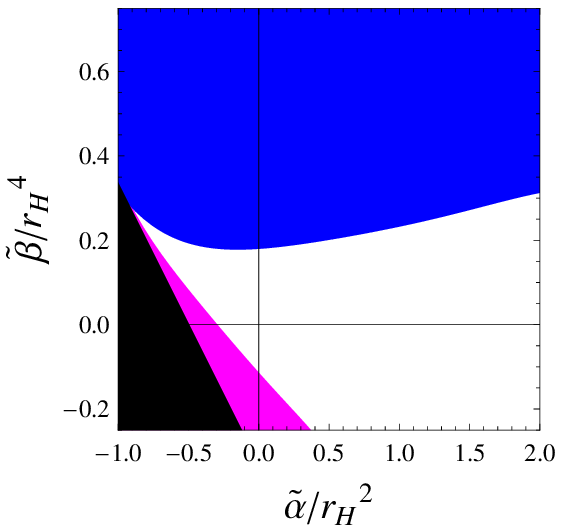}}
\caption{Instabilities for de Sitter black holes ($n=5$) for $r_H/r_C=0$ (flat), $r_H/r_C=0.5$, $r_H/r_C=0.7$, $r_H/r_C=0.9$ (from left to right). Black region is the excluded parametric region, yellow - ghost instability, blue - eikonal instability in the scalar channel, magenta - eikonal instability in the tensor channel, green - eikonal instability in both tensor and scalar channel.}\label{fig:dS}
\end{figure}

\begin{figure}
\resizebox{\linewidth}{!}{\includegraphics*{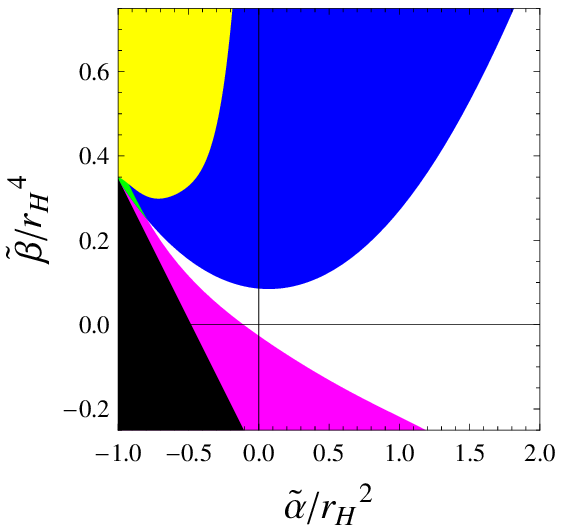}\includegraphics*{n8flat.eps}\includegraphics*{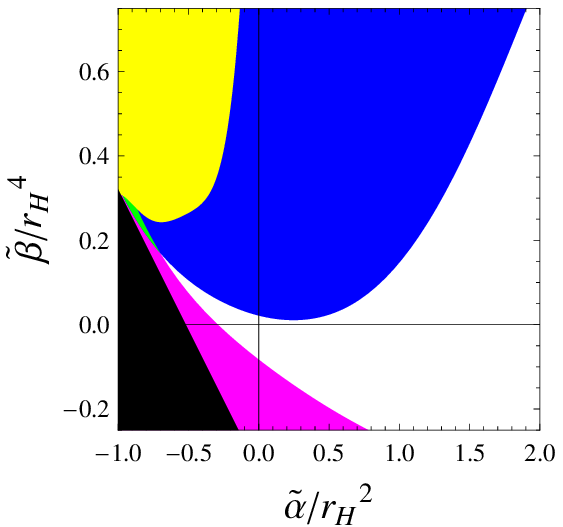}\includegraphics*{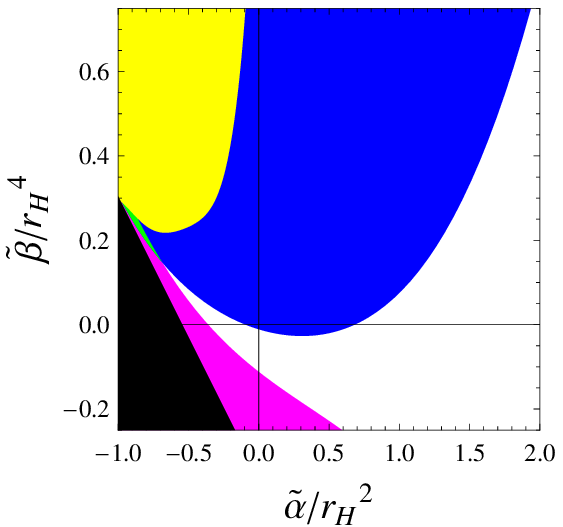}}
\caption{Instabilities for black holes in the flat space ($\Lambda=0$, $n=8$) for $\a_4=-0.01r_H^6$, $\a_4=0$, $\a_4=0.01r_H^6$, $\a_4=0.02r_H^6$ (from left to right). Black region is the excluded parametric region, yellow - ghost instability, blue - eikonal instability in the scalar channel, magenta - eikonal instability in the tensor channel, green - eikonal instability in both tensor and scalar channel.}\label{fig:flat4}
\end{figure}

For asymptotically flat and de Sitter cases, the regions of instability looks different (see figs.~\ref{fig:flat},~\ref{fig:dS},~\ref{fig:flat4}). From fig.~\ref{fig:flat} one can see that the region of stability is increasing at higher $D$. Larger values of the positive $\Lambda$-term also enlarge the stability region and strongly decrease the region of ghost instability, which seems to disappear in the extremal case $r_H = r_C$. From fig.~\ref{fig:flat4} one can see that small values of the third order coupling $\a_4$  deforms the instability region of asymptotically flat black hole only slightly.

\section{Conclusion}\label{sec:conclusion}

While a few particular cases (e.g. at fixed $D$ or asymptotic) of instability for Gauss-Bonnet and Lovelock theories were previously considered in the literature, no one studied the general case of Lovelock theory. Here we performed an exhaustive analysis of eikonal instabilities of black holes in the $D$-dimensional Lovelock theory allowing for the flat, de Sitter and anti-de Sitter asymptotic behavior. The regime of large (in comparison with the AdS radius $R$) black holes naturally includes the case of black branes in AdS spacetime.

We have obtained in a closed form the physically relevant parametric region and the corresponding solutions of the third order Lovelock theory (Sec.~\ref{sec:third-order}).
For the general case, we have chosen several most representative plots to demonstrate the regions of eikonal instability for various $D$, coupling constants, asymptotic behavior etc. We provide here a Mathematica\textregistered{} code which for any given set of parameters says whether this set of parameters describes a black hole having the einsteinian limit and, if so, whether at these values of parameters the black hole suffers from the eikonal instability. In addition, we presented a method of numerical calculation of the line element and the effective potentials for the physically relevant branch of the Lovelock black holes.

\begin{table}
\centering
  \begin{tabular}{|l|l|l|}
      \hline
     black-hole type & eikonal stability & stability at the lowest $\ell$ \\
          \hline
     Schwarzschild  & yes  & yes \cite{Ishibashi:2003ap,Ishibashi:2011ws} \\
          \hline
     Schwarzschild-dS & yes  & yes \cite{Kodama:2003ck,Konoplya:2007jv} \\
          \hline
     Schwarzschild-AdS & yes  & yes \cite{Konoplya:2003dd,Friess:2006kw} \\
          \hline
     Reissner-Nordstr\"om & yes & yes \cite{Kodama:2003ck,Konoplya:2008au} \\
          \hline
     Reissner-Nordstr\"om-dS & yes  & no, for $D\geq7$ \cite{Konoplya:2008au} \\
          \hline
     Reissner-Nordstr\"om-AdS & yes  & yes \cite{Konoplya:2008rq}\\
          \hline
     Schwarzschild-GB & no \cite{Dotti:2005sq,Gleiser:2005ra} & yes \cite{Konoplya:2008ix} \\
      \hline
     Schwarzschild-GB-dS & no \cite{Konoplya:2017ymp,Cuyubamba:2016cug} & no, for $D\geq5$ \cite{Cuyubamba:2016cug}  \\
       \hline
     Schwarzschild-GB-AdS & no \cite{Konoplya:2017ymp,Takahashi:2010gz} & yes \cite{Konoplya:2017ymp} \\
       \hline
     Schwarzschild-Lovelock& no \cite{Takahashi:2010gz} & - \\
       \hline
     Schwarzschild-Lovelock-dS& no & - \\
       \hline
     Schwarzschild-Lovelock-AdS& no & - \\
       \hline
     Reissner-Nordstr\"om-Lovelock& no \cite{Takahashi:2012np,Takahashi:2011qda} & - \\
       \hline
   \end{tabular}
\caption{Review of linear stability of static $D>4$-dimensional black holes in the Einstein and Einstein-Lovelock gravities. Here $D$-dimensional Schwarzschild, Schwarzschild-dS, Reissner-Nordstr\"om etc. metrics mean the corresponding generalizations of the Tangherlini solution \cite{Tangherlini:1963bw}).}\label{Table2}
\end{table}

One should keep in mind that the described here regions of ghost and eikonal instability do not exclude possible instabilities at the lowest multipole number $\ell=2$ (for reviews on black hole stability see \cite{Konoplya:2011qq,Ishibashi:2011ws}). Moreover, such a non-eikonal instability is proved to exist in the Einstein-Gauss-Bonnet-de Sitter case \cite{Cuyubamba:2016cug} and therefore should exist also in a higher order Lovelock theory. Having in mind the both types of instability, let us here briefly review the existing results on the linear stability of static higher dimensional black-holes in the Einstein and Einstein-Lovelock gravities (see Table~\ref{Table2}. While it was analytically shown that $D$-dimensional asymptotically flat Schwarzschild black holes are gravitationally stable \cite{Ishibashi:2003ap}, the cases of non-zero $\Lambda$-term and charge $Q$ required analysis of quasinormal spectra \cite{Konoplya:2007jv,Konoplya:2008rq,Konoplya:2008au}. In particular, in \cite{Konoplya:2008au} it was found that $(D\geq7)$-dimensional Reissner-Nordstr\"om-dS black holes are unstable at the lowest $\ell=2$ multipole. A similar phenomena was observed for the neutral Einstein-Gauss-Bonnet-de Sitter black holes \cite{Cuyubamba:2016cug} for $D\geq5$. In both cases this instability occurs owing to the positive, non-zero $\Lambda$-term and, therefore, it was called $\Lambda$-instability. Summarizing, while eikonal instability exists for all black holes in Gauss-Bonnet and Lovelock theories, the lower-multipole instability seems to take place in addition in the asymptotically de Sitter spacetimes.

\begin{acknowledgments}
The authors thank Andrei Starinets for useful and encouraging discussions. R.~K. would like to thank the Rudolf Peierls Centre for Theoretical Physics of University of Oxford for hospitality and partial support and the Bridging Grant of the University of T\"ubingen. A.~Z. thanks Conselho Nacional de Desenvolvimento Cient\'ifico e Tecnol\'ogico (CNPq) for support and Theoretical Astrophysics of Eberhard Karls University of T\"ubingen for hospitality. At its final stage this work was supported by ``Project for fostering collaboration in science, research and education'' funded by the Moravian-Silesian Region, Czech Republic and by the Research Centre for Theoretical Physics and Astrophysics, Faculty of Philosophy and Science of Sileasian University at Opava.
\end{acknowledgments}

\end{document}